# Evaluation of some Information Retrieval models for Gujarati Ad hoc Monolingual Tasks

**JOSHI Hardik**
Department of Computer Science,
Gujarat University, Ahmedabad

hardikjjoshi@gmail.com

**PAREEK Jyoti**
Department of Computer Science,
Gujarat University, Ahmedabad

drjyotipareek@yahoo.com

**Abstract:** This paper describes the work towards Gujarati Ad hoc Monolingual Retrieval task for widely used Information Retrieval (IR) models. We present an indexing baseline for the Gujarati Language represented by Mean Average Precision (MAP) values. Our objective is to obtain a relative picture of a better IR model for Gujarati Language. Results show that Classical IR models like Term Frequency Inverse Document Frequency (TF_IDF) performs better when compared to few recent probabilistic IR models. The experiments helped to identify the outperforming IR models for Gujarati Language.

## 1 Introduction

Information retrieval (IR) is concerned with representing, searching, and manipulating large collections of electronic text data. IR is the discipline that deals with retrieval of unstructured data or partially structured data, especially textual documents, in response to a set of query or topic statement(s), which may itself be unstructured [6]. The typical interaction between a user and an IR system can be modeled as the user submitting a query to the system; the system returns a ranked list of relevant documents, with the most relevant at top of the list.

The need for effective methods of automated IR has grown in importance because of the tremendous explosion in the amount of text documents and growing number of document sources on the Internet. Over the last few years, there has been a significant growth in the amount of text documents in Indian languages. Researchers have been performing IR tasks in English and European languages since many years [2, 15], efforts are being made to encourage IR tasks for the Indian Languages [5, 14].

Most of the IR research community uses resources known as test collection [12]. Since 1990's, TREC [15] is conducting evaluation exercises using test collections. The classic components of a test collection are:

- A collection of documents; each document is given a unique identifier docid or docno.
- A set of topics (also referred as queries); each query is uniquely identified by a qid or num.
- A set of relevance judgements (also referred as qrels) that consists of a list of (qid,docid) pairs detailing the relevance of documents to topics.

In this paper, we have described the ad hoc monolingual IR task performed over Gujarati language test collection. In ad hoc querying, the user formulates any number of arbitrary queries but applies them to a fixed collection [7]. We have considered Gujarati language, the reason being no such tasks have been performed for Gujarati language, although some work already exists for Bengali, Hindi and Marathi languages [10,13]. Apart from this Gujarati is spoken by nearly 50 Million people over the world and is an official language for the state of Gujarat [9].

The rest of the paper is organized as follows: Section 2 gives an overview of the experimental setup used to perform ad hoc task. Section 3 presents an overview of the evaluation conducted. Section 4 presents the results obtained during the experiment and finally Section 5 concludes the experiments.

## 2. Experimental Setup

### 2.1 Overview of the Corpus

The test collection used for this experiment is the collection made available during the FIRE 2011 [5]. The details of Gujarati Collection are mentioned in Table 1. The collection was created from the archives of the daily newspaper, "Gujarat Samachar" from 2001 to 2010. Each document represents a news article from "Gujarat Samachar". The average number of tokens per document is 445.

| Size of Collection | 2.7 GB |
|---|---|
| Number of text Documents | 3,13,163 |
| Size of Vocabulary | 20,92,619 |
| Number of Tokens | 13,92,72,906 |

Table 1 Statistics of the Gujarati Collection

The corpus is coded in UTF-8 and each article is marked up using the following tags:

<DOC> : Starting of the document
<DOCNO> </DOCNO> : Unique identifier of the document
<TEXT> </TEXT> : Contains the document text
</DOC> : Ending tag of the document

### 2.2 Queries

The IR models were tested against 50 different queries in Gujarati language. Following the TREC model [16], each query is divided into three sections: the title (T) which indicates the brief title, the description (D) that gives a one-sentence description and the narrative part (N), which specifies the relevance assessment criteria. Below is an example of a single query in the collection of 50 queries.

<top>
<num>150</num>
<title>બિલ ગેટ્સ ના પરોપકારી પ્રયત્નો.</title>
<desc>બિલ ગેટ્સનો  માઇક્રોસોફ્ટ થી  નિવૃત્ત થઈને દાનવૃત્તિ કરવાનો નિર્ણય.</desc>
<narr>સંબંધિત દસ્તાવેજો માં માઇક્રોસોફ્ટ ના મુખ્ય બિલ ગેટ્સ તેના પોસ્ટ પરથી નિવૃત્ત થઈને દાન અને સામાજિક કામ કરવાનો નિર્ણય વિષે ની માહિતી હશે.</narr>
</top>

## 2.3 IR Models

In the experiment, we have compared various models that are widely used in test collection evaluation exercise. We considered classical models like Term Frequency Inverse Document Frequency (TF_IDF) model, language models like Hiemstra Language Model (Hiemstra_LM) [8], probabilistic models like Okapi (BM25), Divergence from Randomness (DFR) group of models [1] like Bose-Einstein model for randomness which considers the ratio of two Bernoulli's processes for first Normalization, and Normalization 2 for term frequency normalization (BB2), The DLH hyper-geometric DFR model (DLH) and its improvement (DLH13), Divergence From Independence model (DFI0), A different hyper-geometric DFR model using Popper's normalization (DPH) which is parameter free, DFR based hyper-geometric models which takes an average of two information measures (DFRee), Inverse Term Frequency model with Bernoulli after-effect (IFB2), Inverse Expected Document Frequency model with Bernoulli after-effect (In_expC2), Inverse Document Frequency model with Laplace after-effect (InL2), Poisson model with Laplace after-effect (PL2), a log-logistic DFR model ( LGD) [4] and an Unsupervised DFR model that computed the inner product of Pearson's X^2. In all seventeen different models were evaluated. Few of the models required parametric values; we have used the default values that are generally applied to similar tests.

## 3 Evaluations

In earlier years, the IR systems were evaluated using measures like Precision, Recall and Fallout [3], where precision measures the fraction of retrieved documents that are relevant whereas recall measures the fraction of relevant documents retrieved and fallout measures the fraction of non-relevant documents retrieved. In recent years Mean Average Precision (MAP) values are considered to give the best judgment in the presence of multiple queries [11].

In our experiments, to evaluate the retrieval performance, the mean average precision (MAP) values were considered. This measure is highly recommended among the TREC community and provides a single-figure measure of quality across recall levels. Among evaluation measures, MAP has been shown to have especially good discrimination and stability.

Average Precision is the average of the precision value obtained for the set of top k documents existing after each relevant document is retrieved, and this value is then averaged over multiple queries to obtain MAP. So, if each query qj belongs to a set of queries Q and if the set of relevant documents for a query is {d1, …d_{mj}}and R_{jk} is the set of ranked retrieval results from the top result until we get document d_k then MAP can be calculated using equation 1.

$$\text{MAP (Q)} = \frac{1}{|Q|} \sum_{j=1}^{|Q|} \frac{1}{m_j} \sum_{k=1}^{m_j} \text{Precision (R}_{jk}) \qquad \ldots\ldots\ldots(1)$$

## 4. Results

The MAP values of various IR models are given in Table 2, the top three performers are In_expC2, In_L2, TF_IDF with the highest MAP values. We have evaluated the outperforming IR models with some other evaluation measures like gm_map, Rpref, bpref and reciprocal rank to further compare these models. The results of further investigations are presented in the Table 3.

| IR Model | Mean Average Precision (MAP) Values | | |
|---|---|---|---|
| | T | TD | TDN |
| BB2 | 0.2203 | 0.2371 | 0.0965 |
| BM25 | 0.2181 | 0.2493 | 0.1742 |
| DFI0 | 0.1940 | 0.1267 | 0.0184 |
| DFR_BM25 | 0.2192 | 0.2512 | 0.1747 |
| DFRee | 0.2149 | 0.2189 | 0.1229 |
| DLH | 0.2143 | 0.2314 | 0.1566 |
| DLH13 | 0.2177 | 0.2282 | 0.1449 |
| DPH | 0.2197 | 0.2380 | 0.1525 |
| Hiemstra_LM | 0.2035 | 0.2350 | 0.1806 |
| IFB2 | 0.2156 | 0.1900 | 0.0639 |
| **In_expC2** | 0.2238 | **0.2619** | **0.1989** |
| **InL2** | **0.2240** | 0.2531 | 0.1759 |
| Js_KLs | 0.2183 | 0.2234 | 0.1287 |
| LGD | 0.2186 | 0.2197 | 0.1343 |
| PL2 | 0.1991 | 0.2422 | 0.1468 |
| **TF_IDF** | 0.2198 | 0.2512 | 0.1698 |
| XSqrA_M | 0.2181 | 0.2239 | 0.1309 |

Table 2: MAP values of various IR models

| Measures | IR Models | | |
|---|---|---|---|
| | In_expC2 | InL2 | TF_IDF |
| iprec_at_0.0 | 0.7382 | 0.7198 | **0.7446** |
| Precision@5 | 0.4480 | 0.4960 | **0.5160** |
| gm_map | **0.1894** | 0.1706 | 0.1689 |
| Rpref | **0.3014** | 0.2847 | 0.2892 |
| Bpref | **0.3129** | 0.3072 | 0.3080 |
| recip_rank | 0.6654 | 0.6678 | **0.6971** |

Table 3: Some other Evaluation metrics among top IR Models

The result shown in Table 3 signifies that TF_IDF is outperforming models In_expC2 and InL2 in terms of Precision values. A better picture of TF_IDF model can be obtained from the Graph given in Figure 1, which models the Precision Recall Curve for the TF_IDF Model.

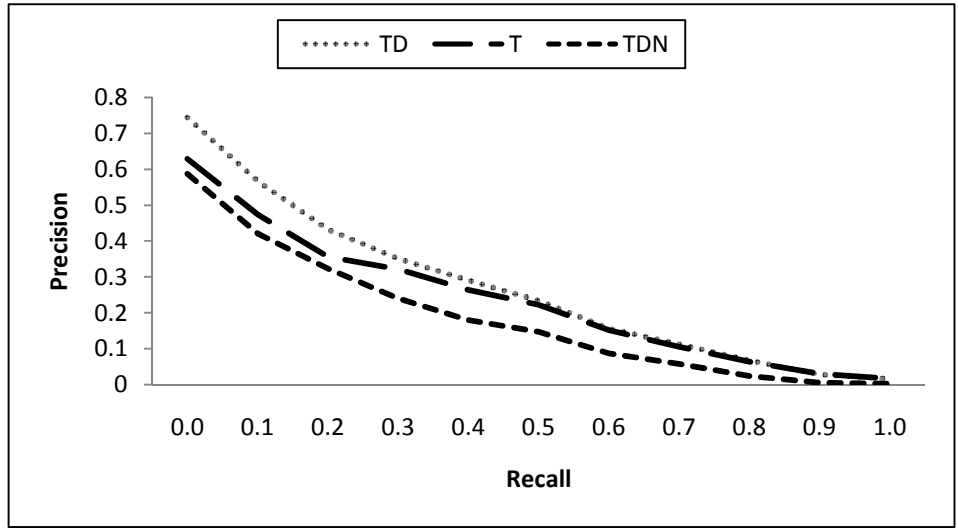

Figure 1: Precision-Recall Curve for the TF_IDF Model

## 5. Conclusion

From the results of experiments carried out to evaluate various IR models, we can conclude that In_expC2, In_L2, and TF_IDF models are performing well with the Gujarati Newswire corpus. Surprisingly TF_IDF is a classical model which does not give good results in English like language. The reason of its performance in Gujarati Language needs to be investigated. Further investigations claim that Precision can be best obtained from the TF_IDF model. Better Precision-Recall curve is obtained when we consider the Title (T) and short Description (D) both together in the queries. Through our experiments, we are able to generate a baseline for each of the IR model for Gujarati Language.